\documentclass[%
 twocolumn,
 superscriptaddress,
 preprintnumbers,
 nofootinbib,
 amsmath,amssymb,
 aps,
 prl
 ]{revtex4-2}
\usepackage{psfrag,slashed,cancel,array,graphicx,hyperref}
\usepackage{subfigure}
\usepackage[utf8]{inputenc}
\usepackage{mathtools}
\usepackage{mathrsfs}
\usepackage{multirow}
\usepackage{braket}
\usepackage{booktabs} %nice tables
\usepackage[normalem]{ulem}
\hypersetup{pdftitle={},pdfcreator={},linkcolor=[rgb]{0.15,0.35,0.75},colorlinks=true,citecolor=[rgb]{0.675,0,0.2},urlcolor=[rgb]{0.15,0.35,0.65}}

\usepackage[dvipsnames]{xcolor}
\newcommand{\colourcolour}[1]{{\color{blue}{#1}}}

\usepackage{hyperref}

\def\dabcnc{{\frac{d^{\,abc}_Fd^{\,abc}_F}{\nF}}}
\def\floo{fl_{11}}
\newcommand{\DD}{{\cal D}}

\def\nc{{n_c}}
\def\ncs{{n_{c}^{\,2}}}
\def\nct{{n_{c}^{\,3}}}
\def\ncf{{n_{c}^{\,4}}}
\def\cI{{C^{}_{\!I}}}
\def\ca{{C_{\!A}}}
\def\cas{{C^{\,2}_A}}
\def\cat{{C^{\,3}_A}}
\def\caf{{C^{\,4}_A}}
\def\cf{{C^{}_{\!F}}}
\def\cfs{{C^{\, 2}_{\!F}}}
\def\cft{{C^{\, 3}_{\!F}}}
\def\cff{{C^{\, 4}_{\!F}}}
\def\nf{{n^{}_{\! f}}}
\def\nfz{{n^{\,0}_{\! f}}}
\def\nfo{{n^{\,1}_{\! f}}}
\def\nfs{{n^{\,2}_{\! f}}}
\def\nft{{n^{\,3}_{\! f}}}

\def\nF{{n_{\!F}^{}}}
\def\nA{{n_{\!A}^{}}}
\def\nI{{n_{\!I}^{}}}

\def\dfIAnI{{\frac{d_I^{\,abcd}d_A^{\,abcd}}{n_I }}}
\def\dfIFnI{{\frac{d_I^{\,abcd}d_F^{\,abcd}}{n_I }}}

\def\dfAAna{{\frac{d_A^{\,abcd}d_A^{\,abcd}}{\nA }}} 
\def\dfFAna{{\frac{d_F^{\,abcd}d_A^{\,abcd}}{\nA }}}
\def\dfFFna{{\frac{d_F^{\,abcd}d_F^{\,abcd}}{\nA }}}

\def\dfFAnc{{\frac{d_F^{\,abcd}d_A^{\,abcd}}{\nF }}}
\def\dfFFnc{{\frac{d_F^{\,abcd}d_F^{\,abcd}}{\nF }}}

\def\dfFA{d^{\,(4)}_{F\!A}}
\def\dfFF{d^{\,(4)}_{F\!F}}

\def\bfctdfFA{{\color{RubineRed}{b_{4,F\!A}^{\rm q}}}}
\def\blackbfctdfFA{{b_{4,F\!A}^{\,\rm q}}}
\def\BBSS{{\mathbb{S}}}
\def\HS{{S}}

\def\frct#1#2{\mbox{\small{$\displaystyle\frac{#1}{#2}$}}}

\newcommand{\pqq}{p_{qq}(x)}
\newcommand{\pqqm}{p_{qq}(-x)}
\newcommand{\HPL}{{\mathrm{H}}}

\newcommand{\als}{\alpha_{\rm s}}
\newcommand{\ars}{a_{\rm s}}

\newcommand{\MSb}{$\overline{\mbox{MS}}$}

\newcommand{\gsim}{\raisebox{-0.07cm}{$\:\stackrel{>}{{\scriptstyle \sim}}\: $} }
\newcommand{\lsim}{\raisebox{-0.07cm}{$\:\stackrel{<}{{\scriptstyle \sim}}\: $} }

\allowdisplaybreaks

\begin{document}

\preprint{DESY 25-067, LTH 1399}

\title{Flavour Non-Singlet Splitting Functions at Four Loops in QCD \\ 
  -- The Fermionic Contributions --
}

\author{B.A.~Kniehl}
\email{bernd.kniehl@desy.de}
\affiliation{II. Institute for Theoretical Physics, Hamburg University, 
 D-22761 Hamburg, Germany} %Luruper Chaussee 149,
\author{S.~Moch}
\email{sven-olaf.moch@desy.de}
\affiliation{II. Institute for Theoretical Physics, Hamburg University, 
 D-22761 Hamburg, Germany} %Luruper Chaussee 149,
\author{V.N.~Velizhanin}
\email{vitaly.velizhanin@desy.de}
\affiliation{II. Institute for Theoretical Physics, Hamburg University, 
 D-22761 Hamburg, Germany} %Luruper Chaussee 149,
\author{A.~Vogt}
\email{andreas.vogt@liverpool.ac.uk}
\affiliation{Department of Mathematical Sciences, University of Liverpool, 
 Liverpool L69 3BX, United Kingdom}

%\date{\today}

%-----------------------------------------------------------------------------%             
\begin{abstract}
We have determined the fourth-order $\nf$ contributions to the two
 splitting functions governing the evolution of all flavor differences of quark 
distributions of hadrons in perturbative quantum chromodynamics with $\nf$ 
light flavors.
The analytic forms of these functions are presented in both Mellin $N$-space
and momentum-fraction $x$-space for a general gauge group.
In the small-$x$ limit double logarithms occur, but the small-$x$ rise of both 
splitting functions is confined to extremely small $x$-values, $x\lsim 10^{-6}$.
The large-$x$ limit includes the $\nf$-part of the four-loop quark virtual 
anomalous dimension. 
Using this result we obtain also the $\nf$ contributions to the corresponding 
gluonic quantity and the complete threshold-enhanced logarithms from soft-gluon
emission for a large class of inclusive observables, including Higgs boson 
production in gluon-gluon fusion.
\end{abstract}
%-----------------------------------------------------------------------------%

% PACS, the Physics and Astronomy Classification Scheme.
% \keywords{QCD}

\maketitle

%-----------------------------------------------------------------------------%
%
Splitting functions are universal quantities in quantum chromodynamics (QCD), 
the gauge theory of the strong interaction.
They describe collinear parton dynamics and govern the scale dependence of 
parton distribution functions (PDFs) of hadrons, playing a crucial role in 
predicting scattering processes with initial-state hadrons.
They are computed in perturbation theory, and are currently fully known at 
three-loop accuracy~\cite{Moch:2004pa,Vogt:2004mw},
corresponding to the next-to-next-to-leading order.
The precision measurements achieved now at the Large Hadron Collider
\cite{Dainese:2019rgk}, as well as anticipated advancements at the 
forthcoming Electron Ion Collider \cite{AbdulKhalek:2021gbh}, 
challenge the existing theoretical accuracy.
This development requires the inclusion of the next quantum loop, 
specifically four loops (N$^3$LO), in QCD evolution equations.

\vspace{0.5mm}
For the non-singlet quark PDFs $q_{\rm ns}^{\pm}(x,\mu^2)$ of hadrons, given 
by flavor differences of quark--anti-quark sums (+) and differences ($-$), 
the evolution equations read$\,$\footnote{
  The total valence distribution evolves with
  $P_{\rm ns}^{\,\rm v}(x) = P_{\rm ns}^{\,-}(x) + P_{\rm ns}^{\:\rm s}(x)$;
  the latter function not considered here. For the current state-of-the-art 
  regarding the evolution of the singlet quark and gluon PDFs, see
  \cite{Falcioni:2023luc,Falcioni:2023vqq,Falcioni:2024xyt,Falcioni:2024qpd}
  and references therein.
}
\begin{equation}
\label{eq:nsEvol}
 \frac{d}{d \ln\mu^2} \, q_{\rm ns}^{\pm}
 \, = \, P_{\rm ns}^{\,\pm}  \otimes q_{\rm ns}^{\pm} 
 \: ,
\end{equation}
with $P_{\rm ns}^{\,\pm}(x,\als) 
 \,=\, \sum_{n=0} \ars^{n+1}\,P_{\rm ns}^{(n),\pm}(x)$
denoting the splitting functions in an expansion in the strong coupling  
$\ars=\alpha_s(\mu^2)/(4\pi)$.
The convolution $\otimes$ is taken in the momentum-fraction variable $x$ and
the anomalous dimensions $\gamma_{\rm ns}^{\,\pm}(N)$ are obtained by a Mellin 
transform, 
\begin{equation}
\label{eq:Mtrf}
  \gamma_{\rm ns}^{(n),\pm}(N) \;=\; 
  - \int_0^1 \!dx\; x^{N-1} P_{\rm ns}^{(n),\pm}(x)
\: , 
\end{equation}
where the relative sign is a standard convention.

\vspace{0.5mm}
Exact results for $P_{\rm ns}^{(3),\pm}$ are available in the large-$\nc$ 
limit of the color $\text{SU}(\nc)$ gauge group~\cite{Moch:2017uml} and for 
the (small) leading and subleading contributions in the limit of a large number 
of flavors $\nf$~\cite{Gracey:1994nn,Davies:2016jie,Basdew-Sharma:2022vya}.
In addition, the $\nf \cft$ contribution is now known~\cite{Gehrmann:2023iah}, 
as well as all terms in $\gamma_{\rm ns}^{(3),\pm}$ which are proportional to 
values of the Riemann $\zeta$-function
\cite{Moch:2017uml,Davies:2017hyl,Kniehl:2025jfs,Moch:2025xxx}.
Beyond these all-$N$ results, the fixed-$N$ values up to $N=22$ have been 
computed for the complete anomalous dimensions $\gamma_{\rm ns}^{(3),\pm}$
\cite{Moch:2017uml,Moch:2025xxx},
together with partial results reaching higher values of $N$.

In this article we provide, for the first time, the complete fermionic 
contributions, i.e. all terms proportional to $\nf$. 
The analytic forms of both $\gamma_{\rm ns}^{(3),\pm}(N)$ and, by virtue of 
eq.~(\ref{eq:Mtrf}), of $P_{\rm ns}^{(3),\pm}(x)$ are presented for a general 
gauge group. 
The determination of all $\nf$ terms in $\gamma_{\rm ns}^{(3),\pm}(N)$ is 
accomplished by reconstructing the entire functional form from fixed-$N$ 
results using LLL-based techniques~\cite{Lenstra:1982eee,fplll} for systems 
of Diophantine equations.  
This all-$N$ analytic reconstruction can be restricted to those parts of the 
anomalous dimensions which are invariant under conformal symmetry.
This significantly reduces the necessary number of fixed-$N$ values.

Our new results have wide-ranging implications due to the universality of soft 
and collinear parton dynamics. 
The $\nf$ terms in $P_{\rm ns}^{(3),\pm}$ enable the analytic determination 
of the quark virtual anomalous dimension at four loops, i.e. the terms 
proportional to $\delta(1-x)$ in $P_{\rm ns}^{(3),\pm}$, except for one 
coefficient known numerically to high accuracy.
Exploiting the known correspondence between collinear gluon radiation off 
quarks and gluons, we also find the gluon virtual anomalous dimension at 
this order.
In addition, due to the universal structure of subleading infrared poles
\cite{Ravindran:2006cg,Dixon:2008gr}, the complete soft corrections at four 
loops for a large class of inclusive observables are now known analytically 
up to the same one numerical coefficient.
These include the coefficient functions for single-inclusive electron-positron
annihilation (SIA) \cite{Moch:2009my,Xu:2024rbt}, 
inclusive deep-inelastic scattering (DIS)~\cite{Das:2019btv}
as well as for the Drell-Yan (DY) process and for Higgs production 
in gluon-gluon fusion (ggF)~\cite{Das:2020adl,Ahmed:2020amh}.

\begin{figure}[bh]
\begin{center}
\vspace*{-2mm}
\includegraphics[width=0.18\textwidth,angle=0]{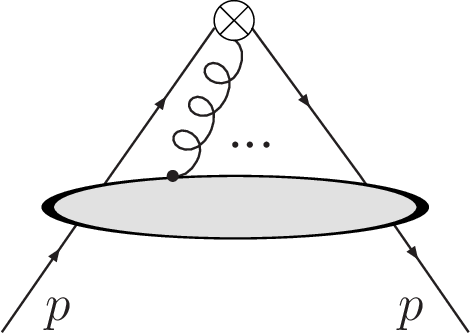}
\vspace*{-2mm}
\caption{\small 
\label{fig:OME} 
Two-point functions with an insertion of the operator
$O^{\,\rm ns}_{\{\,\mu^{\,}_1,\ldots ,\,\mu^{\,}_N\}}$ 
(represented by $\otimes$), computed up to four loops.
}
\vspace*{-7mm}
\end{center}
\end{figure}

Fixed-$N$ values of $\gamma_{\rm ns}^{}$ in eq.~(\ref{eq:Mtrf}) 
-- even-$N$ for $\gamma_{\rm ns}^{\,+}$ and odd-$N$ for 
   $\gamma_{\rm ns}^{\,-}$ --
can be obtained from matrix elements of leading-twist spin-$N$ local 
non-singlet quark operators, 
\begin{eqnarray}
\label{eq:loc-ops}
 O^{\,\rm ns}_{\{\,\mu^{\,}_1, \ldots ,\,\mu^{}_{\!N}\}} &\! =\! & 
   \overline{\psi}\;\lambda^{\alpha}\,\gamma_{\,\{\mu^{}_1}
   D_{\!\mu^{}_2} \ldots D_{\!\mu^{}_{\!N}\}}\,\psi 
   \:\: ,
\end{eqnarray}
where $\psi\, (\overline{\psi})$ are (anti-)quark fields, $D_{\mu^{}_i}$
covariant derivatives and $\lambda^{\alpha}$ the flavor $\text{SU}(\nf)$ 
generators, $\alpha \,=\,3,8, \dots ,$ $ (\nfs-1)$. 
The symmetric and trace-less parts of operators $O^{\,\rm ns}$, 
denoted by $\{\dots\}$ in eq.~(\ref{eq:loc-ops}), are inserted
into quark two-point functions, and the workflow for the computation 
of the operator matrix elements (OMEs), depicted in fig.~\ref{fig:OME}, 
follows~\cite{Moch:2017uml,Moch:2025xxx}.
The Feynman diagrams for the calculation of the OMEs are generated using 
{\bf \texttt{Qgraf}} \cite{Nogueira:1991ex} and then processed by the 
computer algebra system {\bf \texttt{Form}} 
\cite{Vermaseren:2000nd,Kuipers:2012rf,Ruijl:2017dtg} which classifies 
them according to their color factors \cite{vanRitbergen:1998pn} and  
topologies.
The resulting propagator-type integrals for fixed $N$ are computed with 
the {\bf \texttt{Forcer}} package~\cite{Ruijl:2017cxj}. 
In this manner, the complete $\gamma_{\rm ns}^{\,(3),\pm}(N)$ have been 
obtained at $N \leq 22$, and their $\nf$ parts to at least $N = 24$~\cite{Moch:2017uml,Moch:2025xxx}.
The computations are limited by the run-time and the size of intermediate 
expressions in the four-loop integral reductions of the {\bf \texttt{Forcer}} 
program.

The analytic expressions for $\gamma_{\rm ns}^{(3),\pm}(N)$ can be 
written in terms of powers of $N$ and $N\!+\!1$ and harmonic 
sums~\cite{Vermaseren:1998uu}
\begin{equation}
\label{eq:Hsum1}
  S_{\pm m_1^{},\,m_2^{},\,\dots, m_d}(N) \,=\, \sum_{j=1}^{N}\:
  (\pm 1)^{j} \: j^{\, -m_1^{}}\: S_{m_2^{},\,\dots, m_d}(j)
\:\: ,
\end{equation}
of depth $d$ and weight $w = \sum_{i=1}^d |m_i| \leq 7$, where $m_1 > 0$.  
Sums with index -1 do not occur, leaving $1,\, 3,\,7,\,17,\,41,\,99,\dots$ 
sums at $w = 1,\,2,\,3,\,4,\,5,\,6,\dots$, i.e., their number is given 
by the Pell-Lucas numbers \cite{oeis-A001333} which can be evaluated by
$((1 - \sqrt{2})^w + (1 + \sqrt{2})^w)/2$. 
The resulting large size of the space of functions can be dramatically 
reduced by exploiting theoretical constraints on the non-singlet anomalous 
dimensions.
In particular, $\gamma_{\rm ns}^{}$ is expressible through a universal 
evolution kernel $\gamma_{\rm u}^{}$
as~\cite{Dokshitzer:2005bf,Basso:2006nk,Dokshitzer:2006nm},
\begin{equation}
  \label{eq:RR}
  \gamma_{\rm ns}^{}(N) \,=\, \gamma_{\rm u}^{} 
  \left( N - \gamma_{\rm ns}^{}(N)-\beta(\ars)/\ars ) \right)
\:\: ,
\end{equation}
which is reciprocity-respecting (RR), i.e., invariant under the replacement 
$N \to -1\!-\!N$.
Here $\beta(\ars)$ is the standard $\beta$-function in QCD,
known to five loops
\cite{Baikov:2016tgj,Herzog:2017ohr,Luthe:2017ttg,Chetyrkin:2017bjc}.
The perturbative expansion of eq.~(\ref{eq:RR}) separates the
RR part from non-RR terms, so that we have schematically
\begin{equation}
  \label{eq:nth-order-RR}
  \gamma_{\rm ns}^{(n)} \,=\, \gamma_{\rm u}^{(n)} 
  + \text{terms fixed by lower orders}
\:\: .
\end{equation}
A suitable space of RR functions for $\gamma_{\,\rm u}^{(n)}$ is provided by 
powers of $\eta = 1/[N(N\!+\!1)]$ 
-- $\eta^n$ contributes $n$ to the overall weight -- 
and the binomial sums~\cite{Vermaseren:1998uu}
\begin{equation}
\label{eq:BSsum1}
  \BBSS_{m_1^{}, \dots, m_d}(N) = \sum_{j=1}^{N}\,
  (-1)^{N\!+j} \binom{N}{j}\! \binom{N\!\!+\!j}{j} S_{m_1^{}, \dots, m_d}(j)
\: .
\end{equation}
Here all $m_i$ are positive; the number of sums of weight $w$ is $2^{w-1}$. 
At four loops, $2^{w+1}\!-\!1$ functions up to an overall weight $w \leq 7$ can 
contribute to the most complicated (non-$\zeta$) parts of $\gamma_{\rm u}^{}$, 
but considerable fewer enter its $\nf$ contributions. 
This total size of the space of basis functions is much small than that
for $\gamma_{\rm ns}^{(3)}$, which is given by $a(w+2)$ of sequence 
\cite{oeis-A005409}, with, e.g., 127 and 255 instead of 407 and 984 functions 
up to $w=6$ and $w=7$.

\vspace{0.5mm}
A considerable number of coefficients of the basis functions (which are 
integer in a suitable normalization) is further fixed by information on the 
behavior of $\gamma_{\rm ns}^{}$ for $N \!\to\! 0$ and $N \!\to\! \infty$.
In $x$-space, these correspond to the expansions of the splitting functions 
at small-$x$ in terms of double logarithms, $P_{\rm ns}^{(n),\pm}(x) \propto 
\ln^{\,2n-k} x$, where $0 \leq\, k \leq\, 2n-1$~\cite{Vogt:2012gb,Velizhanin:2014dia,Davies:2022ofz}, 
and at large-$x$, where 
$P_{\rm ns}^{(n),\pm}(x) = A_{\rm q}^{(n+1)}\!\!/(1-x)_+ + 
B_{\rm q}^{(n+1)} \delta (1-x) + \dots$ in terms of the cusp anomalous 
dimension $A_{\rm q}^{(\ell)}$, known to $\ell=4$ 
\cite{Henn:2019swt,vonManteuffel:2020vjv},
and the virtual anomalous dimension~$B_{\rm q}$.

\vspace{0.5mm}
The color structure of the $\nf$ part of $\gamma_{\rm ns}^{(3),\pm}$ is 
given by the quadratic (quartic) Casimir invariants $\cf (\dfFF)$ of 
the fundamental representation, and the quadratic adjoint one $\ca$.
%The $\nf \cft$ part has been computed in \cite{Gehrmann:2023iah}, we confirm their result.
We confirm the $\nf \cft$ part computed in \cite{Gehrmann:2023iah}. 
The contributions with $\nf \ca \cfs$, $\nf \cas C_F$ and $\nf \dfFF$ are new.
Only two of the latter are independent, as the large-$\nc$ limit 
of $\gamma_{\rm ns}^{(3),\pm}$ is completely known~\cite{Moch:2017uml}.
We choose to reconstruct the \mbox{all-$N$} form for $\nf \ca \cfs$ from the
moments $N\leq 32$, and that for $\nf \dfFF$ from $N\leq 24$.
Following~\cite{Velizhanin:2010cm} and subsequent works~\cite{Velizhanin:2012nm,Moch:2014sna,%
Davies:2016jie,Moch:2017uml,Falcioni:2023tzp,Kniehl:2025jfs},
we have determined the coefficients of a suitable ansatz by solving a system of
Diophantine equations using the LLL algorithm~\cite{Lenstra:1982eee}
%here specifically the {\bf \texttt{fplll}} package~\cite{fplll}.
as implemented in the {\bf \texttt{fplll}} package~\cite{fplll}.

\vspace{0.5mm}
The $\nf \dfFF$ contribution to $\gamma_{\rm ns}^{(3),\pm}$ is RR by itself, 
since this colour factor does not enter any lower-order terms:
\begin{eqnarray}
\label{eq:gnsdf44}
  \lefteqn{{
  \gamma_{\rm ns}^{(3)\sigma}(N)\Bigr|_{\colourcolour{\nf\*\dfFF}} =
%%START
%%L %%texgnsd4RRnfrat =
  \frct{32}{3}\*\bigl[
  4 \,\* \BBSS_{1,2,1,2} - 4 \,\* \BBSS_{1,3,1,1} - 4 \,\* \BBSS_{2,1,1,2}
%%
%%STOP
  }}
\nonumber
%%START
%%
\\[-1mm] \nonumber 
&&\mbox{}  + 4 \,\* \BBSS_{2,2,1,1} 
 + 12 \* (\BBSS_{1,1,2,1} - \BBSS_{1,2,1,1}) \* \eta \* (1 - \eta) 
 - 6 \,\* \BBSS_{2,2,1} 
\\ \nonumber 
&&\mbox{} - 4 \,\* \BBSS_{2,1,2} \,\* \eta 
 + 2 \,\* \BBSS_{3,1,1} \* (3 + 2 \* \eta) 
 + 3 \,\* \BBSS_{3,1} \* (1 - 4 \* \eta) 
\\ \nonumber 
&&\mbox{} - 12 \,\* \BBSS_{1,2,1} \* \eta \* (1 - \eta^2) 
 + 2 \* \,\BBSS_{2,1,1} \* (5 + 6 \* \eta - 6 \* \eta^3) 
\\ \nonumber 
&&\mbox{} - \BBSS_{2,2} \* (13 - 12 \* \eta) 
 + 2 \,\* \BBSS_{1,2} \* (6 - 2 \* \eta + \eta^2 + 2 \* \eta^3) 
\\ \nonumber 
&&\mbox{} - \BBSS_{2,1} \* (5 - 39 \* \eta) 
 - \BBSS_{2} \* (59 - 28 \* \eta + 5 \* \eta^2 + 6 \* \eta^3) 
\\ \nonumber 
&&\mbox{} + \BBSS_{1,1} \* \eta \* (24 - 39 \* \eta) 
 + \BBSS_{1} \* \eta \* (43 - 85 \* \eta) 
\\ \nonumber 
&&\mbox{}
 + 18 + 47 \* \eta 
 - 62 \* \eta^2
 + \sigma \* \left( 2 \,\* \BBSS_{3} \* \eta \* (3 + \eta + 2 \* \eta^2) \right.
\\ \nonumber
&&\mbox{}
 + 3 \,\* \BBSS_{2} \* \eta \* (1 - 5 \* \eta - 2 \* \eta^2)
 - \eta \* (59 - 62 \* \eta) \left. \right) \bigr]
%%;
%%STOP
\\
&&\mbox{}
   + \zeta_3, \zeta_5\mbox{-terms}
\, ,
\end{eqnarray}
where the arguments $N$ of the $\BBSS$-sums have been omitted.
Remarkably, the few terms of overall weight $w=7$, such as 
$\eta^3\,\BBSS_{2,1,1}$, in eq.~(\ref{eq:gnsdf44}) are exactly those that 
already occurred in the all-$N$ expression in the large-$\nc$ limit
\cite{Moch:2017uml}. 
The complete ansatz for the LLL-reconstruction of the $\nf\*\dfFF$ part 
had 148 coefficients, of which as many as 106 turned out to be zero and 
25 differed by powers of 2. As expected, the highest-weight functions,
with overall weight $w \geq 5$, have particularly simple coefficients.
Together, these observations provide a powerful check of the correctness 
of the all-$N$ result in eq.~(\ref{eq:gnsdf44}). 
 
%\vspace{0.5mm}
The complete expressions for the $\nf$ part of $\gamma_{\rm ns}^{(3),\pm}(N)$ 
can be found in the supplementary material, together with the corresponding
splitting functions $P_{\rm ns}^{(3),\pm}(x)$ expressed in terms of 
harmonic polylogarithms \cite{Remiddi:1999ew,Gehrmann:2001pz}.
%by inverting the Mellin transformation (\ref{eq:Mtrf}). 
% 
The small-$x$ limit of $P_{\rm ns}^{(3),+}(x)$ in terms of $\ln^n\! x$,
$n=1,\dots, 5$ provides more checks of our new exact expressions for the 
$\nf \dfFF$, $\nf \ca \cfs$ and $\nf \cas \cf$ contributions, since none 
of the corresponding coefficients derived in \cite{Velizhanin:2014dia,Davies:2022ofz} 
%, in part using the double logarithmic equation \cite{Velizhanin:2014dia},
were used as constraints in the determinations of the result (\ref{eq:gnsdf44})
and its $\nf \ca \cfs$ counterpart.

%\vspace{0.5mm}
The $\nfo$ parts of $P_{\rm ns}^{(3),+}$ and $P_{\rm ns}^{(3),-}$ 
are shown, in a normalization very close to an expansion in $\als$,
in figs.~\ref{fig:2}(a) and~\ref{fig:2}(b), together with their common large-$\nc$ 
(Lnc) limit and the error bands of \cite{Moch:2017uml} respectively based on 
the first eight even-$N$ and odd-$N$ moments (\ref{eq:Mtrf}). 
The latter provide very good approximations of $P_{\rm ns}^{(3),\pm}$ at 
$x \gsim 10^{-2}$, but cannot sufficiently constraint these functions at much 
smaller values of $x$.
The Lnc limit is very close to both QCD results for $x \gsim 0.3$, and a good
guide to $x \lsim 10^{-3}$.

%\vspace{0.5mm}
Fig.~\ref{fig:2}(c) shows the cumulative effect of the leading 
small-$x$ log (LL), $\ln^5 \!x$, the next-to leading log (NLL) etc for
$P_{\rm ns}^{(3),+}$; the pattern for $P_{\rm ns}^{(3),-}$ is similar.
The LL, NLL, \dots\,  small-$x$ expansions
%csm and their all-order resummations
by themselves are thus of no phenomenological use at any small-$x$ values 
relevant to measurements, a pattern that has been observed before, e.g., 
in \cite{Moch:2004pa,Davies:2022ofz}. 
As shown in fig.~\ref{fig:2}(d), the $\nfo$ parts of 
$P_{\rm ns}^{(3),\sigma}$ exhibit  a small-$x$ rise only below
$x \approx 10^{-6}$ for $\sigma = 1$ and below $x \approx 10^{-7}$ for $\sigma = -1$.
Furthermore, the comparison of the results for an $\text{SU}(2)$ gauge group 
with those for QCD ($\nc=3$) and the Lnc limit of the $\text{SU}(\nc)$ 
illustrates both the size and the quality of convergence of the $1/\nc$ corrections as $\nc$ grows.
Clearly, only exact results in full QCD, like the ones presented here, can reliably 
provide $P_{\rm ns}^{(n),\pm}(x)$ down to values of $x$ relevant to DIS 
off ultra-high energy neutrinos, 
see, e.g.,~\cite{Gauld:2019pgt,Reno:2023sdm}.

\begin{widetext}
\vspace*{-7mm}
\begin{center}
\begin{figure}[!ht]
  \includegraphics[width=0.495\textwidth]{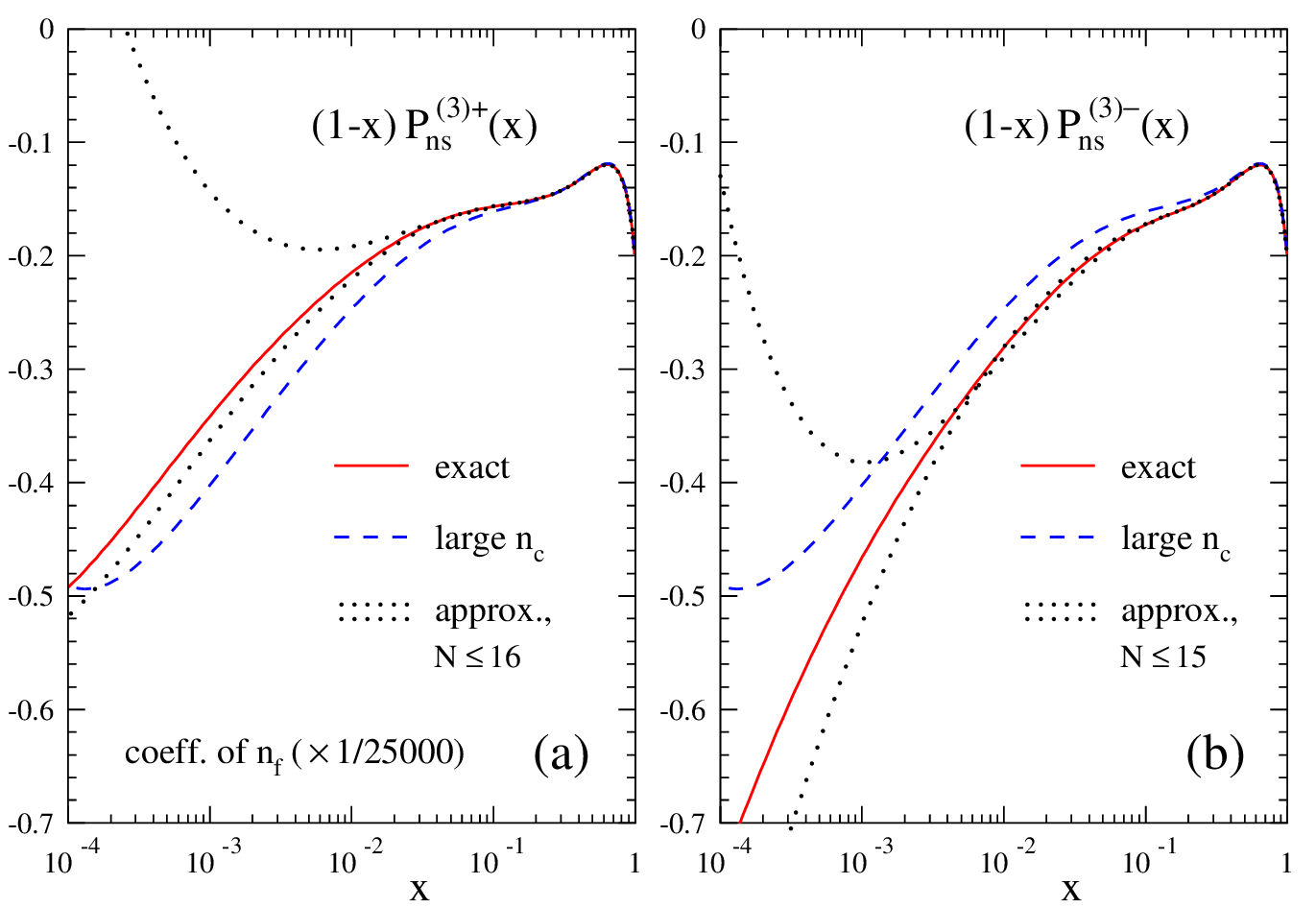}
  \includegraphics[width=0.495\textwidth]{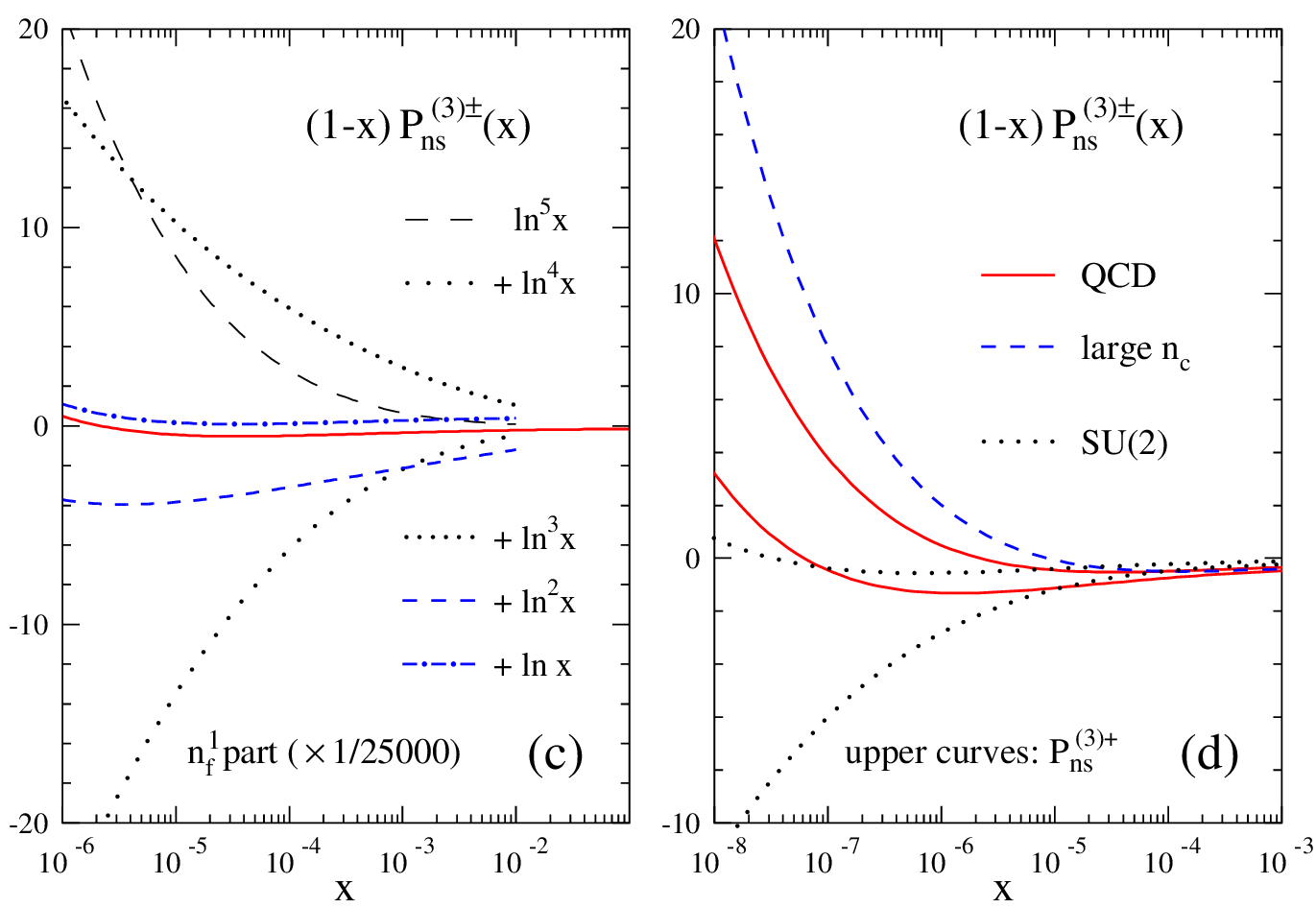}
%\flushleft{
%   \hspace*{21mm} (a)
%   \hspace*{38mm} (b)
%   \hspace*{38mm} (c)
%   \hspace*{38mm} (d)
%}
\vspace*{-6mm}
  \caption{\label{fig:2}
  The exact $\nf$ terms of (a) $P_{\rm ns}^{(3),+}(x)$ and (b) $P_{\rm ns}^{(3),-}(x)$ 
  (always in solid red) compared with their large-$\nc$ limits (dashed blue) 
  and the previous error bands (dotted black) presented in \cite{Moch:2017uml}.
  The exact results down very small values of $x$, 
  compared (c) to the successive `approximations' by the 
%csm  at $x \geq 10^{-6}$ 
  leading, next-next-to-leading, \dots logarithms, and (d)
%csm  at $x \geq 10^{-8}$ 
  to the large-$\nc$ limit (dashed blue) and the results 
  for the gauge group SU(2) instead of SU(3) (dotted black).
  }
\vspace*{-9mm}
\end{figure}
\end{center}
\end{widetext}

The quark virtual anomalous dimension $B_{\rm q}$ can be read off from the
$\delta(1-x)$ terms of $P_{\rm ns}^{\pm}(x)$. We find for QCD
\begin{eqnarray}
\label{eq:B4q-num}
  \nonumber
  B_{\,\rm q}^{(4)} &=& 
  23393.79 \pm 0.02 \,-\, 5550.0446\,\*\nf 
\\[-0.5mm] & & \mbox{}
  \!+\, 193.85504\,\*\nfs \,+\, 3.0149820\,\*\nft
\; ,
\end{eqnarray}
where exact results have been rounded to eight significant digits. 
The complete $\nf$ part of $B_{\rm q}^{(4)}$ at four loops is a new result;
the full-color results are given in the supplementary material.
The corresponding  four-loop gluon virtual anomalous dimension 
$B_{\rm g}^{(4)}$ reads, see also \cite{Moch:2023tdj},
\begin{eqnarray}
\label{eq:B4g-num}
\nonumber
  B_{\rm g}^{(4)} &=& \;\;
  68587.64 \pm 0.3 \,-\, 18143.982\,\*\nf 
  \\[-0.5mm] & & \mbox{}
  \!+\, 423.81135\,\*\nfs \,+\, 0.90672154\,\*\nft
\; .
\end{eqnarray}
This result has been derived with the help of the four-loop quark and gluon 
form factors in QCD~\cite{Henn:2019swt,vonManteuffel:2020vjv}:
their infrared poles are related to the respective virtual anomalous 
dimensions, $B_{\rm q}$ and $B_{\rm g}$, and eikonal anomalous 
dimensions, $f_{\rm q}$ and $f_{\rm g}$, corresponding to color-charged 
Wilson lines~\cite{Dixon:2008gr}.
The latter exhibit the same maximal non-Abelian color structure as the cusp 
anomalous dimensions, thus $f_{\rm q}$ and $f_{\rm g}$ are related by 
(generalized \cite{Moch:2018wjh}) Casimir scaling.
The non-$\nf$ parts of $B_{\rm q}^{(4)}$ and $B_{\rm g}^{(4)}$ are therefore 
also related and share one color coefficient which is not known analytically 
so far. 
In view of the known large-$\nc$ limit, there is some freedom in the 
parametrization of this remaining unknown.
We choose the coefficient of the quartic Casimir invariant $\dfFA$ in 
$B_{\,\rm q}^{(4)}$, for which a precise numerical 
result is available~\cite{Moch:2023tdj,Moch:2025xxx}, 
$\blackbfctdfFA = - 998.02 \pm 0.02$.
This value has been used in eqs.~(\ref{eq:B4q-num}), (\ref{eq:B4g-num})
and the equations below.

\vspace{0.5mm}
The corresponding numerical QCD results for the four-loop eikonal anomalous 
dimensions read
\begin{eqnarray}
\label{eq:f4q-num}
  f_{\rm q}^{\,(4)} &=& \mbox{}
    -\, 22284.01 \pm 0.04 
  \,+\, 3242.924\,\*\nf 
\nonumber \\[-0.5mm] & & \mbox{}
    +\, 26.33248\,\*\nfs
  \,+\, 0.4858066\,\*\nft
  \, ,
\\ 
  f_{\rm g}^{\,(4)} &=& \mbox{}
    -\, 52384.87 \pm 0.6
  \,+\, 7210.446\,\*\nf 
\nonumber \\[-0.5mm] & &  \mbox{}
    +\, 59.24807\,\*\nfs
  \,+\, 1.093065\,\*\nft
\, .
\label{eq:f4g-num}
\end{eqnarray}
The full results are given in the supplementary material.
Other factorizations of soft and collinear parton dynamics,
such as Soft-Collinear Effective Theory (SCET), introduce partonic jet 
functions, see, e.g.,~\cite{Banerjee:2018ozf,Duhr:2022cob,Xu:2024rbt}.
The anomalous dimensions $\gamma_{J,\,\rm I}^{}$ of quark and gluon 
jet functions in SCET are given by 
$\gamma_{J,\,\rm I}^{} = - B_{\,\rm I} - f_{\,\rm I}$ for $I\!=\!{\rm q,g}$ 
and can be read off from eqs.~(\ref{eq:B4q-num})--(\ref{eq:f4g-num}), 
where the numerical uncertainties are anti-correlated in the 
differences.
 
\vspace{0.5mm}
Inclusive cross sections for processes like SIA, DIS, DY or ggF are 
evaluated via partonic coefficient functions, calculable in perturbative QCD. 
They exhibit threshold enhanced double logarithms proportional to 
$\DD_{k}=[\ln^k(1-x)/(1-x)]_+$ in terms of a momentum space variable $x$ 
related to Feynman (or Bjorken) $x$ in case of SIA (DIS), or the ratio 
$x=M^2/S$ of the boson's invariant mass $M$ and the center-of-mass energy 
$\sqrt{S}$ for DY and ggF.
At large $x$, the dominant channels read 
\begin{equation}
\label{eq:cX}
c_{P}^{}(x) = \sum_{n=0}\: a_{\rm s}^n\, c^{(n)}_{P}(x)
 \,\simeq\, \delta(1-x) 
 + \sum_{n=1} \sum_{k=0}^{2n-1} a_{\rm s}^n c^{(n,k)}_{P}\DD_{k}
\:\: ,
\end{equation}
for the projections $P=T, {\rm q}$ for SIA (transverse fragmentation function), 
$P=2, {\rm q}$ for DIS ($F_2$ structure function), 
as well as for quark anti-quark annihilation ($P={\rm q\bar{q}}$) in DY and
ggF Higgs-boson production ($P={\rm gg}$).

\vspace{0.5mm}
The all-order resummation of the threshold logarithms in eq.~(\ref{eq:cX}) is 
currently established to the fourth logarithmic order (N$^{4}$LL),
cf.\ for resummations in SIA~\cite{Moch:2009my,Xu:2024rbt}, 
DIS~\cite{Das:2019btv}, DY and ggF~\cite{Das:2020adl,Ahmed:2020amh}. 
The above virtual and eikonal anomalous dimensions provide the input 
for the determination of the coefficients of $\DD_{0}$ 
in eq.~(\ref{eq:cX}) at four loops.
Eqs.~(\ref{eq:B4q-num})--(\ref{eq:f4g-num}) lead to the new results  
\begin{align}
\label{eq:SIA-cT-numerics}
& c^{{\,\rm SIA}, (4,0)}_{T, {\rm q}} = 
  140708.73 \pm 0.02 
  - 30298.49\, \*\nf 
\nonumber \\[-0.5mm] 
& \;\;
  + 1131.851\, \*\nfs
  - 12.08488\, \*\nft
  + 47.55183\, \*\nf \* \floo 
\, ,
% \end{align}
%%
\\
%% 
%\begin{align}
\label{eq:DIS-c2-numerics}
& c^{{\,\rm DIS}, (4,0)}_{2, {\rm q}} = 
  38840.45 \pm 0.02 
 - 34964.90\, \*\nf 
\nonumber \\[-0.5mm]
& \;\;
  + 2062.715\, \*\nfs
  - 12.08488\, \*\nft
  + 47.55183\, \*\nf \* \floo 
\, ,
%\end{align}
%%
\\
%% 
%\begin{align}
\label{eq:DY-cq-numerics}
& c^{{\,\rm DY}, (4,0)}_{{\rm q{\bar q}}} = 
  - 1279330.4 \pm  0.1
  - 73197.45\, \*\nf 
\nonumber \\[-0.5mm]
& \;\;
  + 10025.71\, \*\nfs
  - 135.1441\, \*\nft
\, ,
%\end{align}
%%
\\
%% 
%\begin{align}
\label{eq:ggF-cg-numerics}
& c^{{\,\rm ggF}, (4,0)}_{{\rm gg}} = 
  - 14987411.0 \pm 1.1 
  - 595317.8\, \*\nf 
\nonumber \\[-0.5mm]
& \;\;
  + 33907.65\, \*\nfs
  - 304.0743\, \*\nft
\, ,
\end{align}
where, in terms of the fractional quark charge 
$e_{\rm q}$, $\floo = 3 \langle e \rangle = (3/\nf) \sum_{\rm q} e_{\rm q}$ 
for flavor non-singlet and 
$\floo = (1/\nf) \langle e \rangle^2 / \langle e^2 \rangle = (1/\nf) 
(\sum_{\rm q^\prime} e_{\rm q^\prime})^2/(\sum_{\rm q} e_{\rm q}^2)$
for singlet coefficient functions~\cite{Vermaseren:2005qc}.
For charged-current DIS one has $\floo=0$. 
The complete results with full color information are given in the 
supplementary material. 

\vspace{0.5mm}
The new $\DD_{0}$ terms in 
eqs.(\ref{eq:SIA-cT-numerics}) -- (\ref{eq:ggF-cg-numerics}) 
also readily allow for an analytic determination, up to the same one color 
coefficient $\blackbfctdfFA$, of the process-dependent resummation 
coefficients governing soft-gluon exponentiation at N$^{4}$LL accuracy for 
SIA, DIS, DY or ggF, 
cf.~\cite{Xu:2024rbt,Das:2019btv,Das:2020adl,Duhr:2022cob} 
for the relevant formulae.
In particular, for Higgs boson production in ggF, this leads to an improved 
perturbative stability of the QCD predictions, i.e., the apparent convergence 
of the perturbative expansion at approximate N$^4$LO accuracy and its stability
under scale variations.
This underscores the importance of the above results, which are largely 
attributable to the universal nature of soft and collinear parton dynamics. 

\vspace{0.5mm}
Exact full-QCD results for also the $\nfz$ parts of the four-loop non-singlet 
splitting functions will require considerable further progress, 
either on their computation via differential equations 
\cite{Basdew-Sharma:2022vya,Gehrmann:2023iah} 
or via the determination of more fixed-$N$ moments for the 
LLL-based reconstructions, supplemented by improved insights into their
analytical structure.
The complete determination of the flavor-singlet cases is even more 
challenging. Already the determination of their $x^{-1} \ln x$ contributions
would suffice, though, in combination with the moments of 
\cite{Falcioni:2023luc,Falcioni:2023vqq,Falcioni:2024xyt,Falcioni:2024qpd},
to reduce their presently remaining numerical uncertainties at small $x$ to a
phenomenologically irrelevant level.

\vspace{2mm}
\begin{acknowledgments}
\emph{Acknowledgments:}  
This work\footnote{
{\bf \texttt{Form}} files with our results can be obtained from the preprint 
server \href{http://arXiv.org}{http://arXiv.org} by downloading the source of 
this article.  They are also available from the authors upon request.}
has been supported by Deutsche Forschungsgemeinschaft (DFG), 
the European Research Council (ERC)
and the UK Science and Technology Facilities Council (STFC).
B.A.K.\ and V.N.V.\ are supported in part through DFG Grants 
KN \mbox{365/13-2} and KN 365/16-1,
S.M. by the ERC Advanced Grant 101095857, {\it Conformal-EIC},
and A.V. by the STFC Consolidated Grant ST/X000699/1.
\end{acknowledgments}

%----------------------------------------------------------------------------%
\cleardoublepage

%\bibliographystyle{apsrev}
%csm \bibliographystyle{JHEP}
%csm \bibliography{main}

\providecommand{\href}[2]{#2}\begingroup\raggedright\endgroup

\newpage
\cleardoublepage
\appendix
%
% ---------------------------------------------------------------------
%
\renewcommand{\theequation}{A.\arabic{equation}}
\setcounter{equation}{0}
\renewcommand{\thefigure}{A.\arabic{figure}}
\setcounter{figure}{0}
\renewcommand{\thetable}{A.\arabic{table}}
\setcounter{table}{0}

\begin{widetext}
\section{Supplemental material}

The quadratic Casimir invariants are $\ca= \nc$ and 
$\cf= (\ncs-1)/(2\nc)$ in $\text{SU}(\nc)$. 
The higher group invariants are products of two symmetrized traces of 
generators $T_r^a$ of the fundamental ($F$) or adjoint ($A$) representations,
\begin{equation}
  d_{F}^{\,abc} \; =\; \frac{1}{2}\: {\rm Tr} \, ( \, 
   T_{F}^{a\,} T_{F}^{b\,} T_{F}^{c\,} 
   + \,\mbox{$bc$ perm.}\, ) 
\; ,
\qquad
  d_{r}^{\,abcd} \; =\; \frac{1}{6}\: {\rm Tr} \, ( \, 
   T_{r}^{a\,} T_{r}^{b\,} T_{r}^{c\,} T_{r}^{d\,}
   + \,\mbox{ five $bcd$ perm.}\, ) 
\; ,
\end{equation}
which leads to
\begin{eqnarray}
\label{eq:d4fxSUn}
\dabcnc &\!=
\displaystyle
  \frac{( \ncs -4 )\,( \ncs - 1 )}{16\,\ncs} 
\;\,, \quad
\dfFAnc &\!=
  \frac{( \ncs -1 )\,( \ncs + 6 )}{48} 
\;\,, \quad
\dfFFnc =
  \frac{(\ncs - 1)\,( \ncf - 6\,\ncs + 18 )}{96\,\nct}
  \; ,
  \nonumber \\ 
\dfAAna &\!=
\displaystyle
  \frac{1}{24}\: \ncs ( \ncs + 36 ) 
\;\,, \quad~~~
\dfFAna &\!=
  \frac{1}{48}\: \nc ( \ncs + 6 )
\;\,, \qquad~~
\dfFFna =
\frac{( \ncf - 6\,\ncs + 18 )}{96\,\ncs}
\;\,,
\end{eqnarray}
where $\nF=n_c$ and $\nA=n_c^2-1$.
Their values are
$\dabcnc\,=\, 5/18$, 
$\dfFAnc \,=\, 5/2$, $\dfFFnc \,=\, 5/36$, 
$\dfAAna = 135/8$, $\dfFAna = 15/16$ and $\dfFFna = 5/96$ in QCD.

\vspace*{-1mm}
\subsection{Non-singlet anomalous dimensions and splitting functions}
\vspace*{-2mm}

The integer-$N$ moments (\ref{eq:Mtrf}) of the fermionic contributions to the 
non-singlet four-loop splitting functions $P_{\rm ns}^{(3),\pm}(x)$ in 
eq.~(\ref{eq:nsEvol}) can be expressed in terms of harmonic sums as defined 
in eq.~(\ref{eq:Hsum1}).
We omit the argument $N$ of these sums and use the abbreviations 
$\eta = 1/[(N\!+\!1)N]$, $D_1 = 1/(\!N+\!1)$. 
In the standard \MSb\ scheme adopted throughout this article,
the $\nfo$ even-$N$ parts of $\gamma_{\rm ns}^{(3),+}(N)$ and the 
corresponding odd-$N$ results for $\gamma_{\rm ns}^{(3),-}(x)$ read

The corresponding expressions for the $\nfs$ and $\nft$ contributions can be
found in eqs.~(3.1)--(3.3) and (3.6) of \cite{Davies:2016jie}, see also
eq.~(18) of \cite{Gracey:1994nn}. 
See section 3 of \cite{Moch:2004pa} for the results up to three loops.

The fermionic parts of the non-singlet splitting functions 
$P_{\,\rm ns}^{\,(3),\pm}$ are expressed in terms of harmonic 
polylogarithms~\cite{Remiddi:1999ew}, which we have numerically evaluated with
a weight-6 extension of the routine of \cite{Gehrmann:2001pz}. 
Below, we use
\begin{equation}
\label{eq:habbr}
  H_{{\footnotesize \underbrace{0,\ldots ,0}_{\scriptstyle m} },\,
  \pm 1,\, {\footnotesize \underbrace{0,\ldots ,0}_{\scriptstyle n} },
  \, \pm 1,\, \ldots}(x) \:\equiv\: H_{\pm (m+1),\,\pm (n+1),\, \ldots}(x)
  \:\: .
\end{equation}
Suppressing their argument $x$ and using $\pqq = 2\,(1-x)^{-1} - 1 - x$, 
we find 

where all divergences for $x\!\rightarrow\! 1$ are to be read as plus-distributions. 
See, e.g., \cite{Moch:2004pa} for more details on the regularization of the $x\!\rightarrow\! 1$ limit.
The $\nf \cft$ terms in eqs.~(\ref{eq:Pns3p}) and (\ref{eq:Pns3m}) have already 
been computed in~\cite{Gehrmann:2023iah} and are included here for completeness.
The terms proportional to $\nfs$ and $\nft$ can be found in eqs.~(4.5)--(4.10)
of \cite{Davies:2016jie}, for the lower-order non-singlet splitting functions, 
see section 4 of \cite{Moch:2004pa}.

The small-$x$ expansions of the $\nfo$ parts of the non-singlet splitting 
functions $P_{\,\rm ns}^{\,(3),\pm}$ in eqs.~(\ref{eq:Pns3p}) and 
\ref{eq:Pns3m}) read
\begin{eqnarray}
\label{eq:Pns3p-xto0}
  \nonumber
  {\lefteqn{
\lim_{x \to 0} P_{\,\rm ns}^{\,(3),+}(x)\Bigr|_{\nf} \! =\! }}
\\ && \nonumber
%%START
%%L %%texPns3pnfxto0 = 
          - \frac{4}{9} \*\cft \* \ln^5(x) 
       + \biggl(
          - \frac{20}{9}\*\cft
          - \frac{44}{9}\*\cfs\*\ca
          \biggr) \* \ln^4(x)
       + \biggl(
          - \frac{16}{3}\*\cft
          - \frac{2092}{27}\*\cfs\*\ca
          - \frac{242}{27}\*\cf\*\cas
          + 64\*\zeta_2\*\cft
          - 48\*\zeta_2\*\cfs\*\ca
\\
&& \nonumber
          + 16\*\zeta_2\*\cf\*\cas
           \biggr) \* \ln^3(x)
       + \biggl(
          - \frac{170}{3}\*\cft
          - \frac{32968}{81}\*\cfs\*\ca
          - \frac{1390}{9}\*\cf\*\cas
          + \frac{1840}{3}\*\zeta_2\*\cft
          + \frac{620}{3}\*\zeta_2\*\cf\*\cas
          - \frac{4928}{9}\*\zeta_2\*\cfs\*\ca
\\
&& \nonumber
          + \frac{256}{3}\*\zeta_3\*\cft
          - 64\*\zeta_3\*\cfs\*\ca
           \biggr) \* \ln^2(x)
       + \biggl (
            \frac{500}{3}\*\cft
          - \frac{90538}{81}\*\cfs\*\ca
          - \frac{64481}{81}\*\cf\*\cas
          + \frac{3592}{3}\*\zeta_2\*\cft
          + \frac{6104}{9}\*\zeta_2\*\cf\*\cas
\\
&& \nonumber
          - \frac{9584}{9}\*\zeta_2\*\cfs\*\ca
          + \frac{2080}{3}\*\zeta_3\*\cft
          - \frac{1328}{3}\*\zeta_3\*\cfs\*\ca
          - \frac{32}{3}\*\zeta_3\*\cf\*\cas
          + 304\*\zeta_4\*\cft
          - \frac{1448}{3}\*\zeta_4\*\cfs\*\ca
          + 56\*\zeta_4\*\cf\*\cas
           \biggr) \* \ln(x)
%%;
%%STOP
\\ & & + \: O(1) 
\end{eqnarray}
and
\begin{eqnarray}
\label{eq:dPns3pm-xto0}
  \nonumber
  {\lefteqn{
\lim_{x \to 0} \Bigl(P_{\,\rm ns}^{\,(3),-}(x) - P_{\,\rm ns}^{\,(3),+}(x)\Bigr)\Bigr|_{\nf} \! =\! }}
\\ && \nonumber
%%START
%%L %%texdPns3mpnfxto0 = 
         \frac{8}{15}\*\cf\*(\ca-2\*\cf)^2 \* \ln^5(x)
       + \frac{2}{9}\*\cf\*(\ca-2\*\cf)\* (13\*\ca - 54\*\cf)\* \ln^4(x)
       - \frac{80}{81}\*\cf\*(\ca-2\*\cf)\* (44\*\ca + 63\*\cf + 18\*\zeta_2\*\ca
\\
&& \nonumber
       - 54\*\zeta_2\*\cf)\* \ln^3(x)
       - \frac{4}{81}\*\cf\*(\ca-2\*\cf)\* (3800\*\ca + 9639\*\cf + 5490\*\zeta_2\*\ca - 12096\*\zeta_2\*\cf + 2052\*\zeta_3\*\ca - 1944\*\zeta_3\*\cf)\* \ln^2(x)
\\
&& \nonumber
   + \biggl( 
            2752\*\cft
          - \frac{161200}{81}\*\cfs\*\ca
          + \frac{27104}{81}\*\cf\*\cas
          - 2176\*\zeta_2\*\cft
          + \frac{71168}{27}\*\zeta_2\*\cfs\*\ca
          - \frac{21016}{27}\*\zeta_2\*\cf\*\cas
          - \frac{5824}{3}\*\zeta_3\*\cft
\\
&& \nonumber
          + \frac{22016}{9}\*\zeta_3\*\cfs\*\ca
          - \frac{6712}{9}\*\zeta_3\*\cf\*\cas
          - 720\*\zeta_4\*\cft
          + \frac{2120}{3}\*\zeta_4\*\cfs\*\ca
          - \frac{1624}{9}\*\zeta_4\*\cf\*\cas
          - \frac{3968}{3}\*\dfFFnc
\\
&& 
          + \frac{640}{3}\*\zeta_2\*\dfFFnc
          + 384\*\zeta_3\*\dfFFnc
          + \frac{1024}{3}\*\zeta_4\*\dfFFnc
          \biggr) \* \ln(x)
%%;
%%STOP
   \: + \: O(1) \: .
\end{eqnarray}
The $\zeta_2$ terms in eq.~(\ref{eq:Pns3p-xto0}) are new, as the
double logarithmic equation of \cite{Velizhanin:2014dia} does 
not cover these contributions beyond the large-$\nc$ limit.
All other parts of eq.~(\ref{eq:Pns3p-xto0}) provide checks
of the result in eq.~(\ref{eq:Pns3p}). Eq.~(\ref{eq:dPns3pm-xto0})
is entirely new.

\vspace*{-2mm}
\subsection{Virtual and eikonal anomalous dimensions}
\vspace*{-2mm}
The virtual four-loop quark and gluon anomalous dimensions 
$B_{\rm q}^{(4)}$ and $B_{\rm g}^{(4)}$ are given by
\begin{eqnarray}
\label{eq:Bq4}
  \nonumber
  {\lefteqn{
  B_{\rm q}^{(4)} =\! }}
%%START
%%L %%texBq4 =
\\
&& \nonumber
   \colourcolour{\cff} \* \left( \frac{4873}{24} - 450\*\zeta_2 - \frac{684}{5}\*\zeta_2^2 - \frac{16888}{35}\*\zeta_2^3 + 2004\*\zeta_3 
 - 120\*\zeta_3\*\zeta_2 + \frac{128}{5}\*\zeta_3\*\zeta_2^2 - 1152\*\zeta_3^2
   - 2520\*\zeta_5 - 384\*\zeta_5\*\zeta_2 + 5880\*\zeta_7  \right)
  \\ \nonumber
      &&  + \colourcolour{\cf\*\cat} \* \left( -\frac{371201}{648} - \frac{1}{24}\* \bfctdfFA + \frac{4582}{3}\*\zeta_2 - \frac{22388}{135}\*\zeta_2^2 
 + \frac{48368}{315}\*\zeta_2^3 - \frac{153670}{81}\*\zeta_3 + \frac{472}{3}\*\zeta_3\*\zeta_2 + \frac{16}{5}\*\zeta_3\*\zeta_2^2 
 + 528\*\zeta_3^2
\right.
\\
&& \nonumber
\left.
         + \frac{11372}{9}\*\zeta_5 + 504\*\zeta_5\*\zeta_2 - 2870\*\zeta_7 \right) 
   + \colourcolour{\cfs\*\cas} \* \left( \frac{29639}{36} - \frac{46771}{27}\*\zeta_2 - \frac{24340}{27}\*\zeta_2^2 - \frac{21988}{35}\*\zeta_2^3 
   + \frac{129662}{27}\*\zeta_3 + \frac{2096}{9}\*\zeta_3\*\zeta_2
\right.
\\
&& \nonumber
\left.
   - \frac{64}{5}\*\zeta_3\*\zeta_2^2 - \frac{7102}{3}\*\zeta_3^2 + \frac{5354}{9}\*\zeta_5 
   - 2104\*\zeta_5\*\zeta_2 + 8610\*\zeta_7 \right)
   + \colourcolour{\cft\*\ca} \* \left( -\frac{2085}{4} + 1167\*\zeta_2 + \frac{4334}{5}\*\zeta_2^2 + \frac{317188}{315}\*\zeta_2^3 
\right.
\\
&& \nonumber
\left.
   - 3260\*\zeta_3
   - \frac{1988}{3}\*\zeta_3\*\zeta_2 + \frac{256}{5}\*\zeta_3\*\zeta_2^2 + 3220\*\zeta_3^2 - 976\*\zeta_5 
   + 2064\*\zeta_5\*\zeta_2 - 10920\*\zeta_7 \right) 
   + \colourcolour{\dfFAnc} \* \left( \bfctdfFA \right)
  \\ \nonumber
      &&  + \colourcolour{\nf\*\cft} \* \left( 32 + 162\*\zeta_2 - \frac{408}{5}\*\zeta_2^2 - \frac{51472}{315}\*\zeta_2^3 - 308\*\zeta_3 
 - \frac{256}{3}\*\zeta_3\*\zeta_2 + 224\*\zeta_3^2 + 912\*\zeta_5 \right) \\ \nonumber
      &&  + \colourcolour{\nf\*\cfs\*\ca} \* \left( -\frac{7751}{54} - \frac{3892}{27}\*\zeta_2 + \frac{55708}{135}\*\zeta_2^2 + \frac{2808}{35}\*\zeta_2^3 
 - \frac{15400}{27}\*\zeta_3 + \frac{2672}{9}\*\zeta_3\*\zeta_2 - \frac{1232}{3}\*\zeta_3^2 - \frac{7432}{9}\*\zeta_5 \right) \\ \nonumber
      &&  + \colourcolour{\nf\*\cf\*\cas} \* \left( \frac{20027}{108} - \frac{41092}{81}\*\zeta_2 + \frac{2468}{45}\*\zeta_2^2 - \frac{4472}{135}\*\zeta_2^3 
 + \frac{9554}{27}\*\zeta_3 - \frac{580}{3}\*\zeta_3\*\zeta_2 + \frac{416}{3}\*\zeta_3^2 + \frac{1130}{9}\*\zeta_5 \right) \\ \nonumber
      &&  + \colourcolour{\nf\*\dfFFnc} \* \left( -192 + \frac{1888}{3}\*\zeta_2 - \frac{704}{15}\*\zeta_2^2 + \frac{2048}{45}\*\zeta_2^3 
 - \frac{992}{3}\*\zeta_3 + 64\*\zeta_3\*\zeta_2 + 256\*\zeta_3^2 - 1120\*\zeta_5 \right) \\ \nonumber
      &&  + \colourcolour{\nfs\*\cfs} \* \left( -\frac{188}{27} + \frac{1244}{27}\*\zeta_2 - \frac{4208}{135}\*\zeta_2^2 + \frac{56}{27}\*\zeta_3 
 - \frac{160}{9}\*\zeta_3\*\zeta_2 + \frac{368}{9}\*\zeta_5 \right)
         + \colourcolour{\nfs\*\cf\*\ca} \* \left( -\frac{193}{54} + \frac{3170}{81}\*\zeta_2 - \frac{32}{9}\*\zeta_2^2
\right.
\\
&& 
\left.
 - \frac{320}{9}\*\zeta_3 + \frac{80}{3}\*\zeta_3\*\zeta_2 
 - \frac{88}{9}\*\zeta_5 \right) 
  + \colourcolour{\nft\*\cf} \* \left( -\frac{131}{81} + \frac{32}{81}\*\zeta_2 - \frac{64}{135}\*\zeta_2^2 + \frac{304}{81}\*\zeta_3 \right)
%%;
%%STOP
\: , \\[2mm] \nonumber
%\end{eqnarray}
%and 
%%
%\begin{eqnarray}
\label{eq:Bg4}
  \nonumber
  {\lefteqn{
  B_{\,\rm g}^{(4)} =\! }}
%%START
%%L %%texBg4 =
\\
&& \nonumber
  \colourcolour{\caf} \* \left( \frac{50387}{486} - \frac{1}{24}\* \bfctdfFA + \frac{2098}{27}\*\zeta_2 + \frac{1793}{27}\*\zeta_2^2 
       - \frac{76516}{945}\*\zeta_2^3 + \frac{48088}{27}\*\zeta_3 - \frac{3902}{9}\*\zeta_3\*\zeta_2 + \frac{336}{5}\*\zeta_3\*\zeta_2^2 + \frac{682}{3}\*\zeta_3^2 
   - \frac{14617}{9}\*\zeta_5
\right.
\\
&& \nonumber
   + 80\*\zeta_5\*\zeta_2 + 700\*\zeta_7 \biggr)
   + \colourcolour{\dfAAna} \* \left( -\frac{800}{9} +  \bfctdfFA + \frac{1184}{3}\*\zeta_2 - \frac{1016}{15}\*\zeta_2^2 
       + \frac{5984}{315}\*\zeta_2^3 - \frac{784}{3}\*\zeta_3 - 272\*\zeta_3\*\zeta_2 + \frac{760}{3}\*\zeta_5 \right) \\ \nonumber
      &&  + \colourcolour{\nf\*\cat} \* \left( -7 - \frac{2579}{27}\*\zeta_2 - \frac{1601}{135}\*\zeta_2^2 + \frac{6908}{945}\*\zeta_2^3 
- \frac{12749}{27}\*\zeta_3 + \frac{718}{9}\*\zeta_3\*\zeta_2 - \frac{836}{3}\*\zeta_3^2 + \frac{931}{3}\*\zeta_5 \right) \\ \nonumber
      &&  + \colourcolour{\nf\*\cf\*\cas} \* \left( -\frac{22627}{486} + \frac{34}{3}\*\zeta_2 + \frac{26}{3}\*\zeta_2^2 + \frac{64}{9}\*\zeta_2^3 
- \frac{2182}{9}\*\zeta_3 - 8\*\zeta_3\*\zeta_2 + 232\*\zeta_3^2 - 80\*\zeta_5 \right)
    + \colourcolour{\nf\*\cfs\*\ca} \* \left( -\frac{1859}{27} + \frac{176}{9}\*\zeta_3 \right) \\ \nonumber
  &&  + \colourcolour{\nf\*\cft} \* ( 23 )
     + \colourcolour{\nf\*\dfFAna} \* \left( \frac{224}{9} - 160\*\zeta_2 + \frac{1328}{15}\*\zeta_2^2 + \frac{2368}{315}\*\zeta_2^3 
 + \frac{320}{3}\*\zeta_3 + 608\*\zeta_3\*\zeta_2 + 256\*\zeta_3^2 - \frac{4880}{3}\*\zeta_5 \right) \\ \nonumber
      &&  + \colourcolour{\nfs\*\cas} \* \left( \frac{1352}{81} + \frac{37}{27}\*\zeta_2 + \frac{80}{27}\*\zeta_2^2 + \frac{289}{27}\*\zeta_3 
         - \frac{32}{9}\*\zeta_3\*\zeta_2 - \frac{8}{9}\*\zeta_5 \right)
   + \colourcolour{\nfs\*\cf\*\ca} \* \left( \frac{3910}{243} + \frac{160}{9}\*\zeta_3 \right) \\ 
      &&  + \colourcolour{\nfs\*\cfs} \* \left( \frac{338}{27} - \frac{176}{9}\*\zeta_3 \right) 
       + \colourcolour{\nfs\*\dfFFna} \* \left( -\frac{704}{9} + \frac{512}{3}\*\zeta_3 \right) 
       + \colourcolour{\nft\*\ca} \* \left( \frac{5}{243} \right) 
       + \colourcolour{\nft\*\cf} \* \left( \frac{154}{243} \right)
%%;
%%STOP
\end{eqnarray}
with, for now, $\blackbfctdfFA = - 998.02 \pm 0.02$ 
\cite{Moch:2023tdj,Moch:2025xxx}.
The corresponding lower-order coefficients can be read off from the 
$\delta(1\!-\!x)$ terms in eqs.~(4.5), (4.6) and (4.9) of \cite{Moch:2004pa}
and eqs.~(4.6), (4.10) and (4.15) of \cite{Vogt:2004mw}.
 
\vspace{1mm}
The eikonal anomalous dimensions for quarks and gluons, $f_{\rm q}^{}$ and 
$f_{\rm g}^{}$, are related by (generalized) Casimir scaling.
For $f_{\rm q}^{(4)}$ we have
$\cI= \cf$, $d_I^{\,abcd} = d_F^{\,abcd}$ and $\nI = \nF$,
and for $f_{\rm g}^{(4)}$ accordingly 
$\cI= \ca$, $d_I^{\,abcd} = d_A^{\,abcd}$ and $\nI = \nA$:
\begin{eqnarray}
\label{eq:fI4}
  \nonumber
  {\lefteqn{
  f_{\,\rm I}^{(4)}  =\! }}
%%START
%%L %%texfI4 =
\\
&& \nonumber
  \colourcolour{\cI\*\cat} \* \left( \frac{9311591}{6561} + \frac{1}{12}\,\*\bfctdfFA - \frac{1164703}{729}\*\zeta_2 + \frac{231518}{135}\*\zeta_2^2 - \frac{173888}{315}\*\zeta_2^3 
 - \frac{829204}{243}\*\zeta_3 + \frac{12568}{9}\*\zeta_3\*\zeta_2 - \frac{4228}{15}\*\zeta_3\*\zeta_2^2 \right.
\\
&& \nonumber
\left.
- \frac{4378}{9}\*\zeta_3^2 + \frac{106934}{27}\*\zeta_5 
 - \frac{1376}{3}\*\zeta_5\*\zeta_2 - \frac{11071}{6}\*\zeta_7 \right) 
  + \colourcolour{\dfIAnI} \* \left( 192 - 2\,\*  \bfctdfFA - \frac{2176}{3}\*\zeta_2 + \frac{224}{15}\*\zeta_2^2 + \frac{27808}{315}\*\zeta_2^3 
\right.
\\
&& \nonumber
\left.
   - \frac{7808}{9}\*\zeta_3 - 1792\*\zeta_3\*\zeta_2 - \frac{736}{5}\*\zeta_3\*\zeta_2^2 - \frac{3344}{3}\*\zeta_3^2 - \frac{1840}{9}\*\zeta_5 
   + 1024\*\zeta_5\*\zeta_2 + 3484\*\zeta_7 \right)
   + \colourcolour{\nf\*\cI\*\cfs} \* \left( \frac{21037}{108} - 2\*\zeta_2 + \frac{148}{5}\*\zeta_2^2 \right.
\\
&& \nonumber
\left.
- \frac{320}{7}\*\zeta_2^3 + \frac{4424}{9}\*\zeta_3 
 - 80\*\zeta_3^2 - \frac{1600}{3}\*\zeta_5 \right) 
   + \colourcolour{\nf\*\cI\*\cf\*\ca} \* \left( -\frac{813475}{972} + \frac{2819}{9}\*\zeta_2 - \frac{1976}{45}\*\zeta_2^2 + \frac{128}{35}\*\zeta_2^3 
+ \frac{68882}{81}\*\zeta_3 \right.
\\
&& \nonumber
\left.
   - 160\*\zeta_3\*\zeta_2 - 312\*\zeta_3^2 + \frac{1448}{9}\*\zeta_5 \right) 
   + \colourcolour{\nf\*\cI\*\cas} \* \left( -\frac{394109}{1944} + \frac{294539}{729}\*\zeta_2 - \frac{4420}{9}\*\zeta_2^2 + \frac{27032}{189}\*\zeta_2^3 
 - \frac{31340}{243}\*\zeta_3 \right.
\\
&& \nonumber
\left.
   - \frac{104}{9}\*\zeta_3\*\zeta_2 + \frac{4420}{9}\*\zeta_3^2 - \frac{692}{27}\*\zeta_5 \right) 
   + \colourcolour{\nf\*\dfIFnI} \* \left( 256\*\zeta_2 - \frac{64}{5}\*\zeta_2^2 - \frac{1280}{21}\*\zeta_2^3 + \frac{640}{9}\*\zeta_3 - \frac{320}{3}\*\zeta_3^2 - \frac{1600}{9}\*\zeta_5 \right) \\ \nonumber
      &&  + \colourcolour{\nfs\*\cI\*\cf} \* \left( \frac{16733}{486} - \frac{172}{9}\*\zeta_2 + \frac{128}{45}\*\zeta_2^2 - \frac{4568}{81}\*\zeta_3 + \frac{32}{3}\*\zeta_3\*\zeta_2 + \frac{304}{9}\*\zeta_5 \right) 
     + \colourcolour{\nfs\*\cI\*\ca} \* \left( \frac{27875}{17496} - \frac{15481}{729}\*\zeta_2 + \frac{776}{45}\*\zeta_2^2 \right.
\\
&& 
\left.
   + \frac{32152}{243}\*\zeta_3 
  - \frac{224}{9}\*\zeta_3\*\zeta_2 - 112\*\zeta_5 \right) 
  + \colourcolour{\nft\*\cI} \* \left( -\frac{16160}{6561} - \frac{16}{81}\*\zeta_2 + \frac{256}{135}\*\zeta_2^2 - \frac{400}{243}\*\zeta_3 \right)
%%;
%%STOP
   \: .
\end{eqnarray}
See eq.~(3.22) in \cite{Das:2019btv} for the lower orders.

\vspace{-2mm}
\subsection{Coefficient functions at large {\bf $x$}}
\vspace{-2mm}
The $\DD_{0}$ term $c^{\,{\rm DIS}, (4,0)}_{2, {\rm q}}$ in the four-loop DIS 
quark coefficient function is given below.
For the terms proportional to $\DD_{k}$ with $1 \leq k \leq 7$, and
the lower-order coefficient functions near threshold,
see eqs.~(A.1)--(A.5) in \cite{Das:2019btv}.
\begin{eqnarray}
\label{eq:c2q40}
  \nonumber
  {\lefteqn{
c^{{\rm DIS}, (4,0)}_{2, {\rm q}}  =\! }}
%%START
%%L %%texc2q40 =
\\ \nonumber
  &&
     \colourcolour{\cff} \* \left(
     \frac{37069}{48} - 2040\*\zeta_7 - 4224\*\zeta_5 - 940\*\zeta_3 + 1888\*\zeta_3^2 + 3253\*\zeta_2 - 4608\*\zeta_2\*\zeta_5 - 5048\*\zeta_2\*\zeta_3 + \frac{20523}{5}\*\zeta_2^2 - 176\*\zeta_2^2\*\zeta_3
\right.
\\
&& \nonumber
\left.
+ \frac{138784}{105}\*\zeta_2^3
     \right)
  + \colourcolour{\cft\*\ca} \* \left( - \frac{492041}{108} + 10920\*\zeta_7 + \frac{32168}{3}\*\zeta_5 + \frac{37640}{3}\*\zeta_3 - \frac{21668}{3}\*\zeta_3^2 - 17850\*\zeta_2 - 2256\*\zeta_2\*\zeta_5
\right.
\\
&& \nonumber
\left.
+ \frac{50296}{9}\*\zeta_2\*\zeta_3 - \frac{441829}{54}\*\zeta_2^2 + \frac{7736}{5}\*\zeta_2^2\*\zeta_3 + \frac{39272}{21}\*\zeta_2^3 
\right) 
  + \colourcolour{\cfs\*\cas} \* \left( \frac{51498031}{3888} - 8610\*\zeta_7 + \frac{14894}{9}\*\zeta_5 - \frac{188162}{27}\*\zeta_3
\right.
\\
&& \nonumber
\left.
+ \frac{5966}{3}\*\zeta_3^2 + \frac{17325892}{729}\*\zeta_2 + 3960\*\zeta_2\*\zeta_5 - \frac{462688}{27}\*\zeta_2\*\zeta_3 - \frac{7041313}{810}\*\zeta_2^2 + \frac{25736}{15}\*\zeta_2^2\*\zeta_3 + \frac{40576}{45}\*\zeta_2^3 
 \right)
\\ \nonumber
  &&
   + \colourcolour{\cf\*\cat} \* \left( -\frac{59835979}{4374} - \frac{1}{24}\,\*\bfctdfFA + \frac{28291}{6}\*\zeta_7 - \frac{128510}{27}\*\zeta_5 + \frac{387083}{27}\*\zeta_3 - \frac{17182}{9}\*\zeta_3^2 + \frac{2052595}{243}\*\zeta_2 - \frac{136}{3}\*\zeta_2\*\zeta_5 
\right.
\\
&& \nonumber
\left.
- \frac{39988}{9}\*\zeta_2\*\zeta_3 - \frac{77341}{45}\*\zeta_2^2 + \frac{836}{3}\*\zeta_2^2\*\zeta_3 + \frac{55864}{189}\*\zeta_2^3 
     \right) 
   + \colourcolour{\dfFAnc} \* \left( - 192 + \bfctdfFA
     - 3484\*\zeta_7 + \frac{1840}{9}\*\zeta_5 + \frac{7808}{9}\*\zeta_3 
\right.
\\
&& \nonumber
\left.
   + \frac{3344}{3}\*\zeta_3^2 + \frac{2176}{3}\*\zeta_2 
   - 1024\*\zeta_2\*\zeta_5 + 1792\*\zeta_2\*\zeta_3 - \frac{224}{15}\*\zeta_2^2 + \frac{736}{5}\*\zeta_2^2\*\zeta_3 - \frac{27808}{315}\*\zeta_2^3 
     \right)
\\ \nonumber
  &&
  + \colourcolour{\nf\*\cft} \* \left( - \frac{16067}{27} - \frac{3952}{3}\*\zeta_5 - \frac{15314}{9}\*\zeta_3 - \frac{496}{3}\*\zeta_3^2 + \frac{33958}{27}\*\zeta_2 + \frac{10432}{9}\*\zeta_2\*\zeta_3 + \frac{38906}{27}\*\zeta_2^2 - \frac{608}{7}\*\zeta_2^3 
 \right)
\\ \nonumber
  &&
  + \colourcolour{\nf\*\cfs\*\ca} \* \left( - \frac{255235}{324} + \frac{2984}{9}\*\zeta_5 - \frac{71482}{27}\*\zeta_3 + \frac{2920}{3}\*\zeta_3^2 - \frac{5369177}{729}\*\zeta_2 + \frac{21952}{9}\*\zeta_2\*\zeta_3 + \frac{915377}{405}\*\zeta_2^2 - \frac{41488}{315}\*\zeta_2^3 
 \right)
\\ \nonumber
  &&
  + \colourcolour{\nf\*\cf\*\cas} \* \left( \frac{31645735}{5832} - \frac{6562}{27}\*\zeta_5 - \frac{239062}{81}\*\zeta_3 - \frac{2612}{9}\*\zeta_3^2 - \frac{940424}{243}\*\zeta_2 + \frac{8684}{9}\*\zeta_2\*\zeta_3 + \frac{3314}{5}\*\zeta_2^2 - \frac{86176}{945}\*\zeta_2^3  \right)
\\ \nonumber
  &&
  + \colourcolour{\nf\*\dfFFnc} \* \left(
     192 + \frac{11680}{9}\*\zeta_5 + \frac{2336}{9}\*\zeta_3 - \frac{448}{3}\*\zeta_3^2 - \frac{2656}{3}\*\zeta_2 - 64\*\zeta_2\*\zeta_3 + \frac{896}{15}\*\zeta_2^2 + \frac{4864}{315}\*\zeta_2^3 
     \right)
\\ \nonumber
  &&
  + \colourcolour{\nfs\*\cfs} \* \left( - \frac{161929}{972} - \frac{64}{9}\*\zeta_5 + \frac{3812}{9}\*\zeta_3 + \frac{385300}{729}\*\zeta_2 - \frac{1376}{27}\*\zeta_2\*\zeta_3 - \frac{19904}{135}\*\zeta_2^2 
 \right)
   + \colourcolour{\nfs\*\cf\*\ca} \* \left( - \frac{3761509}{5832} + \frac{1192}{9}\*\zeta_5 
\right.
\\
&& \nonumber
\left.
   + \frac{6092}{81}\*\zeta_3 + \frac{131878}{243}\*\zeta_2 - \frac{400}{9}\*\zeta_2\*\zeta_3 - \frac{616}{9}\*\zeta_2^2 \right) 
   + \colourcolour{\nft\*\cf} \* \left( \frac{50558}{2187} + \frac{80}{81}\*\zeta_3 - \frac{1880}{81}\*\zeta_2 + \frac{16}{9}\*\zeta_2^2
   \right)
\\ 
  &&
  + \colourcolour{\floo\*\nf\*\cf\*\dabcnc} \* \left( - 192 + 1280\*\zeta_5 - 224\*\zeta_3 - 480\*\zeta_2 + \frac{96}{5}\*\zeta_2^2 \right)
%%;
%%STOP
   \:\: .
\end{eqnarray}
The large-$x$ difference between the SIA and DIS quark coefficient functions 
at four loops has been derived in eq.~(32) of~\cite{Moch:2009my}. 
This includes the $\DD_{0}$ terms 
$c^{{\,\rm SIA}, (4,0)}_{T, {\rm q}} - c^{\,{\rm DIS}, (4,0)}_{2, {\rm q}}$, 
therefore $c^{{\,\rm SIA}, (4,0)}_{T, {\rm q}}$ can be obtained from 
eq.~(\ref{eq:c2q40}).
See eqs.~(7)--(9) in~\cite{Moch:2009my} for the lower-order SIA coefficient 
functions near threshold.

\vspace{1mm}
The $\DD_{0}$ coefficient $c^{\,{\rm DY}, (4,0)}_{{\rm q\bar{q}}}$ of the 
four-loop coefficient function for Drell-Yan quark anti-quark annihilation 
reads
\begin{eqnarray}
\label{eq:cqq40}
  \nonumber
%  {\lefteqn{
  && c^{{\rm DY}, (4,0)}_{{\rm q\bar{q}}} \: =\: 
% }}
% \\ \nonumber %   &&
%%START
%%L %%texcqq40 =
   \colourcolour{\cff} \* \left( 983040\*\zeta_7 - 196608\*\zeta_5 + 32704\*\zeta_3 - 15360\*\zeta_3^2 - 491520\*\zeta_2\*\zeta_5 + 113152\*\zeta_2\*\zeta_3 - \frac{391168}{5}\*\zeta_2^2\*\zeta_3 \right)
\\ \nonumber
  &&  + \colourcolour{\cft\*\ca} \* \left(  - \frac{206444}{27} + 274432\*\zeta_5 - \frac{746878}{9}\*\zeta_3 - \frac{484192}{3}\*\zeta_3^2 - \frac{32740}{9}\*\zeta_2 - 73728\*\zeta_2\*\zeta_5 - \frac{1011088}{9}\*\zeta_2\*\zeta_3 + \frac{293536}{45}\*\zeta_2^2 
\right.
\\
&& \nonumber
\left.
+ 30400\*\zeta_2^2\*\zeta_3 + \frac{406912}{15}\*\zeta_2^3  \right) 
   + \colourcolour{\cfs\*\cas} \* \left(  \frac{15086188}{729} + \frac{1046528}{9}\*\zeta_5 + \frac{3043898}{81}\*\zeta_3 - \frac{82592}{3}\*\zeta_3^2 - \frac{12535492}{729}\*\zeta_2 + 3072\*\zeta_2\*\zeta_5 
\right.
\\
&& \nonumber
\left.
- \frac{2968640}{27}\*\zeta_2\*\zeta_3+ \frac{1008832}{81}\*\zeta_2^2 + \frac{121888}{15}\*\zeta_2^2\*\zeta_3 - \frac{34496}{15}\*\zeta_2^3 \right)
   + \colourcolour{\cf\*\cat} \* \left(  - \frac{28290079}{2187} - \frac{1}{6}\,\* \bfctdfFA
   + \frac{11071}{3}\*\zeta_7 - \frac{149980}{27}\*\zeta_5 
\right.
\\
&& \nonumber
\left.
+ \frac{288544}{9}\*\zeta_3 - \frac{14828}{9}\*\zeta_3^2 + \frac{5746982}{243}\*\zeta_2 + \frac{2752}{3}\*\zeta_2\*\zeta_5 - \frac{120968}{9}\*\zeta_2\*\zeta_3 - \frac{301208}{45}\*\zeta_2^2 + \frac{8456}{15}\*\zeta_2^2\*\zeta_3 + \frac{1009888}{945}\*\zeta_2^3 \right) 
\\
&& \nonumber
   + \colourcolour{\dfFAnc} \* \left( - 384 + 4\,\* \bfctdfFA - 6968\*\zeta_7 + \frac{3680}{9}\*\zeta_5 + \frac{15616}{9}\*\zeta_3 + \frac{6688}{3}\*\zeta_3^2 
   + \frac{4352}{3}\*\zeta_2 - 2048\*\zeta_2\*\zeta_5 + 3584\*\zeta_2\*\zeta_3
\right.
\\
&& \nonumber
\left.
 - \frac{448}{15}\*\zeta_2^2    + \frac{1472}{5}\*\zeta_2^2\*\zeta_3 - \frac{55616}{315}\*\zeta_2^3       \right)
   + \colourcolour{\nf\*\cft} \* \left(  - \frac{73309}{54} - \frac{119680}{3}\*\zeta_5 + \frac{84848}{9}\*\zeta_3 + \frac{109024}{3}\*\zeta_3^2 - \frac{8248}{27}\*\zeta_2 + \frac{186752}{9}\*\zeta_2\*\zeta_3 
\right.
\\
&& \nonumber
\left.
- \frac{37672}{45}\*\zeta_2^2 - \frac{497536}{105}\*\zeta_2^3 \right)
  + \colourcolour{\nf\*\cfs\*\ca} \* \left(  - \frac{1792393}{1458} - \frac{368272}{9}\*\zeta_5 - \frac{1407448}{81}\*\zeta_3 + \frac{6800}{3}\*\zeta_3^2 + \frac{2636774}{729}\*\zeta_2 + 35232\*\zeta_2\*\zeta_3 
\right.
\\
&& \nonumber
\left.
- \frac{370768}{81}\*\zeta_2^2 + \frac{32384}{105}\*\zeta_2^3 \right)
  + \colourcolour{\nf\*\cf\*\cas} \* \left( \frac{11551831}{2916} - \frac{7064}{27}\*\zeta_5 - \frac{829304}{81}\*\zeta_3 - \frac{4552}{9}\*\zeta_3^2 - \frac{2400868}{243}\*\zeta_2 + \frac{23440}{9}\*\zeta_2\*\zeta_3 
\right.
\\
&& \nonumber
\left.
+ \frac{108896}{45}\*\zeta_2^2 - \frac{5872}{21}\*\zeta_2^3 \right)
 + \colourcolour{\nf\*\dfFFnc} \* \left( \frac{3200}{9}\*\zeta_5 - \frac{1280}{9}\*\zeta_3 + \frac{640}{3}\*\zeta_3^2 - 512\*\zeta_2 + \frac{128}{5}\*\zeta_2^2 + \frac{2560}{21}\*\zeta_2^3 
     \right)
\\ \nonumber
  &&  + \colourcolour{\nfs\*\cfs} \* \left(  - \frac{142769}{729} + \frac{33056}{9}\*\zeta_5 + \frac{113456}{81}\*\zeta_3 - \frac{99184}{729}\*\zeta_2 - \frac{79360}{27}\*\zeta_2\*\zeta_3 + \frac{9280}{27}\*\zeta_2^2  \right)
   + \colourcolour{\nfs\*\cf\*\ca} \* \left(  - \frac{898033}{2916} + \frac{608}{3}\*\zeta_5 
\right.
\\
&&
\left.
+ \frac{87280}{81}\*\zeta_3 + \frac{293528}{243}\*\zeta_2 - \frac{608}{9}\*\zeta_2\*\zeta_3 - \frac{3488}{15}\*\zeta_2^2  \right)    + \colourcolour{\nft\*\cf} \* \left( \frac{10432}{2187} - \frac{3680}{81}\*\zeta_3 - \frac{3200}{81}\*\zeta_2 + \frac{224}{45}\*\zeta_2^2
   \right)
%%;
%%STOP
   \:\: .
\end{eqnarray}
The coefficients of $\DD_{k}$ for $2 \leq k \leq 7$ at four loops can be found 
in eq.~(6) of~\cite{Ravindran:2006cg}, and that of $\DD_{1}$ in eq.~(14) 
of~\cite{Ahmed:2014cla} (only in `v1' at {\tt arXiv.org}). 
The third-order $\DD_{k}$ coefficients near threshold were presented in 
eqs.~(17)--(26) of~\cite{Moch:2005ky}, and the $\delta(1\!-\!x)$ term in 
eq.~(12) of \cite{Ahmed:2014cla}.

The $\DD_{0}$ term $c^{\,{\rm ggF}, (4,0)}_{{\rm gg}}$ of the four-loop
coefficient function for Higgs-boson production via gluon-gluon fusion is 
\begin{eqnarray}
\label{eq:cgg40}
  \nonumber
  {\lefteqn{
c^{\,{\rm ggF}, (4,0)}_{{\rm gg}} \! =\! }}
%%START
%%L %%texcgg40 =
\\ \nonumber
  &&
   \colourcolour{\caf} \* \left( - \frac{40463407}{2187} - \frac{1}{6}\,\* \bfctdfFA + \frac{2960191}{3}\*\zeta_7 + \frac{10233380}{27}\*\zeta_5 + \frac{10989496}{81}\*\zeta_3 - \frac{1969484}{9}\*\zeta_3^2 + \frac{28569362}{729}\*\zeta_2 
\right.
\\
&& \nonumber
\left.
- \frac{1683776}{3}\*\zeta_2\*\zeta_5 - \frac{7256984}{27}\*\zeta_2\*\zeta_3 + \frac{5197928}{405}\*\zeta_2^2 - 39144\*\zeta_2^2\*\zeta_3 + \frac{24472096}{945}\*\zeta_2^3 
     \right)
  + \colourcolour{\dfAAna} \* \left( - 384 + 4\,\*\bfctdfFA 
\right.
\\
&& \nonumber
\left.
- 6968\*\zeta_7 + \frac{3680}{9}\*\zeta_5 + \frac{15616}{9}\*\zeta_3 + \frac{6688}{3}\*\zeta_3^2 + \frac{4352}{3}\*\zeta_2 - 2048\*\zeta_2\*\zeta_5 + 3584\*\zeta_2\*\zeta_3 - \frac{448}{15}\*\zeta_2^2 + \frac{1472}{5}\*\zeta_2^2\*\zeta_3 - \frac{55616}{315}\*\zeta_2^3 
 \right)
\\ \nonumber
  &&  + \colourcolour{\nf\*\cat} \* \left( \frac{18455767}{2916} - \frac{2194328}{27}\*\zeta_5 - \frac{3115000}{81}\*\zeta_3 + \frac{269048}{9}\*\zeta_3^2 - \frac{10816300}{729}\*\zeta_2 + 57360\*\zeta_2\*\zeta_3 - \frac{1029536}{405}\*\zeta_2^2 - \frac{514096}{105}\*\zeta_2^3 
 \right)
\\ \nonumber
  &&  + \colourcolour{\nf\*\cf\*\cas} \* \left( \frac{2519645}{486} - \frac{2608}{3}\*\zeta_5 - \frac{143036}{9}\*\zeta_3 + 9712\*\zeta_3^2 - \frac{134542}{27}\*\zeta_2 + \frac{27808}{9}\*\zeta_2\*\zeta_3 - \frac{1424}{5}\*\zeta_2^2 + \frac{3328}{35}\*\zeta_2^3  \right)
\\ \nonumber
  &&  + \colourcolour{\nf\*\cfs\*\ca} \* \left( - \frac{21037}{54} + \frac{3200}{3}\*\zeta_5 - \frac{8848}{9}\*\zeta_3 + 160\*\zeta_3^2 + 16\*\zeta_2 - \frac{296}{5}\*\zeta_2^2 + \frac{640}{7}\*\zeta_2^3 \right)
  + \colourcolour{\nf\*\dfFAna} \* \left( \frac{3200}{9}\*\zeta_5 - \frac{1280}{9}\*\zeta_3 
\right.
\\
&& \nonumber
\left.
+ \frac{640}{3}\*\zeta_3^2 - 512\*\zeta_2 + \frac{128}{5}\*\zeta_2^2 + \frac{2560}{21}\*\zeta_2^3 \right)
+ \colourcolour{\nfs\*\cas} \* \left( - \frac{1543153}{2916} + \frac{34592}{9}\*\zeta_5 + \frac{176624}{81}\*\zeta_3 + \frac{1171400}{729}\*\zeta_2 - \frac{71200}{27}\*\zeta_2\*\zeta_3 
\right.
\\
&& \nonumber
\left.
+ \frac{2336}{45}\*\zeta_2^2  \right)
   + \colourcolour{\nfs\*\cf\*\ca} \* \left( - \frac{155083}{243} + 32\*\zeta_5 + \frac{4784}{9}\*\zeta_3 + \frac{5600}{9}\*\zeta_2 - \frac{1280}{3}\*\zeta_2\*\zeta_3 + \frac{64}{3}\*\zeta_2^2 
 \right) 
\\
  &&   + \colourcolour{\nft\*\ca} \* \left( \frac{10432}{2187} - \frac{3680}{81}\*\zeta_3 - \frac{3200}{81}\*\zeta_2 + \frac{224}{45}\*\zeta_2^2 
   \right)
%%;
%%STOP
   \: .
\end{eqnarray}
The fourth-order coefficients of $\DD_{k}$ for $2 \leq k \leq 7$ have been 
given in eq.~(16) of~\cite{Ravindran:2006cg}, and that of $\DD_{1}$ in 
eq.~(13) of~\cite{Ahmed:2014cla} (only in `v1' at {\tt arXiv.org}). 
The third-order $\DD_{k}$ coefficients can be found in eqs.~(17)--(22) 
of~\cite{Moch:2005ky}. The corresponding $\delta(1\!-\!x)$ term was 
determined in eq.~(10) of \cite{Anastasiou:2014vaa}.

\end{widetext}

\end{document}